\documentclass[twoside,twocolumn,9pt]{article}
\usepackage{extsizes}
\usepackage[super,sort&compress,comma]{natbib} 
\usepackage[version=3]{mhchem}
\usepackage[left=1.5cm, right=1.5cm, top=1.785cm, bottom=2.0cm]{geometry}
\usepackage{balance}
\usepackage{mathptmx}
\usepackage{sectsty}
\usepackage{graphicx} 
\usepackage{lastpage}
\usepackage[format=plain,justification=justified,singlelinecheck=false,font={stretch=1.125,small,sf},labelfont=bf,labelsep=space]{caption}
\usepackage{float}
\usepackage{fancyhdr}
\usepackage{fnpos}
\usepackage[english]{babel}
\addto{\captionsenglish}{%
  }
\usepackage{array}
\usepackage{droidsans}
\usepackage{charter}
\usepackage[T1]{fontenc}
\usepackage[usenames,dvipsnames]{xcolor}
\usepackage{setspace}
\usepackage[compact]{titlesec}
\usepackage{hyperref}
\usepackage{siunitx}
\usepackage{upgreek}
\usepackage{siunitx}
\usepackage{amssymb}
\DeclareSIUnit\angstrom{\text {Å}}
\DeclareSIUnit\bar{bar}

\usepackage{epstopdf}

\definecolor{cream}{RGB}{222,217,201}

\begin{document}

\pagestyle{fancy}
\thispagestyle{plain}
\fancypagestyle{plain}{
\renewcommand{\headrulewidth}{0pt}
}

\makeFNbottom
\makeatletter
\renewcommand\LARGE{\@setfontsize\LARGE{15pt}{17}}
\renewcommand\Large{\@setfontsize\Large{12pt}{14}}
\renewcommand\large{\@setfontsize\large{10pt}{12}}
\renewcommand\footnotesize{\@setfontsize\footnotesize{7pt}{10}}
\makeatother

\renewcommand{\thefootnote}{\fnsymbol{footnote}}
\renewcommand\footnoterule{\vspace*{1pt}%
\color{cream}\hrule width 3.5in height 0.4pt \color{black}\vspace*{5pt}} 
\setcounter{secnumdepth}{5}

\makeatletter 
\renewcommand\@biblabel[1]{#1}            
\renewcommand\@makefntext[1]%
{\noindent\makebox[0pt][r]{\@thefnmark\,}#1}
\makeatother 
\renewcommand{\figurename}{\small{Fig.}~}
\sectionfont{\sffamily\Large}
\subsectionfont{\normalsize}
\subsubsectionfont{\bf}
\setstretch{1.125} 
\setlength{\skip\footins}{0.8cm}
\setlength{\footnotesep}{0.25cm}
\setlength{\jot}{10pt}
\titlespacing*{\section}{0pt}{4pt}{4pt}
\titlespacing*{\subsection}{0pt}{15pt}{1pt}

\fancyfoot{}
\fancyfoot[LO,RE]{\vspace{-7.1pt}\includegraphics[height=9pt]{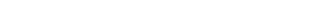}}
\fancyfoot[CO]{\vspace{-7.1pt}\hspace{13.2cm}\includegraphics{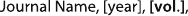}}
\fancyfoot[CE]{\vspace{-7.2pt}\hspace{-14.2cm}\includegraphics{head_foot/RF}}
\fancyfoot[RO]{\footnotesize{\sffamily{1--\pageref{LastPage} ~\textbar  \hspace{2pt}\thepage}}}
\fancyfoot[LE]{\footnotesize{\sffamily{\thepage~\textbar\hspace{3.45cm} 1--\pageref{LastPage}}}}
\fancyhead{}
\renewcommand{\headrulewidth}{0pt} 
\renewcommand{\footrulewidth}{0pt}
\setlength{\arrayrulewidth}{1pt}
\setlength{\columnsep}{6.5mm}
\setlength\bibsep{1pt}

\makeatletter 
\newlength{\figrulesep} 
\setlength{\figrulesep}{0.5\textfloatsep} 

\newcommand{\topfigrule}{\vspace*{-1pt}%
\noindent{\color{cream}\rule[-\figrulesep]{\columnwidth}{1.5pt}} }

\newcommand{\botfigrule}{\vspace*{-2pt}%
\noindent{\color{cream}\rule[\figrulesep]{\columnwidth}{1.5pt}} }

\newcommand{\dblfigrule}{\vspace*{-1pt}%
\noindent{\color{cream}\rule[-\figrulesep]{\textwidth}{1.5pt}} }

\makeatother

\twocolumn[
  \begin{@twocolumnfalse}
{\includegraphics[height=30pt]{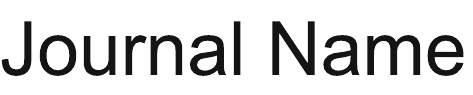}\hfill\raisebox{0pt}[0pt][0pt]{\includegraphics[height=55pt]{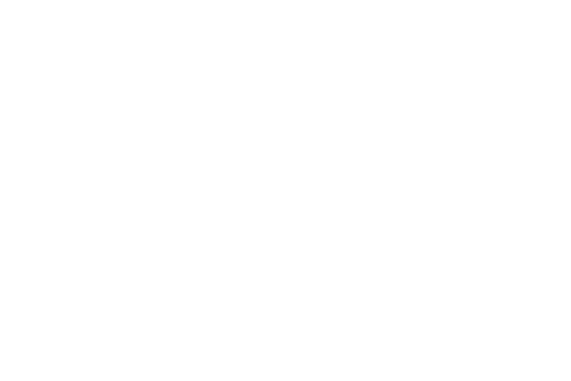}}\\[1ex]
\includegraphics[width=18.5cm]{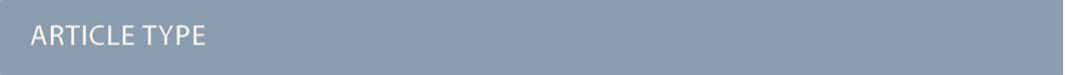}}\par
\vspace{1em}
\sffamily
\begin{tabular}{m{4.5cm} p{13.5cm} }

\includegraphics{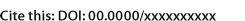} & \noindent\LARGE{\textbf{Growth of Large Crystals of Janus Phase RhSeCl Using Self-Selecting Vapour Growth$^\dag$}} \\
\vspace{0.3cm} & \vspace{0.3cm} \\

& \noindent\large{Anastasiia Lukovkina,$^{\ddag}$\textit{$^{a}$} Maria A. Herz,$^{\ddag}$\textit{$^{a}$}  Xiaohanwen Lin\textit{$^{a}$}, Volodymyr Multian\textit{$^{a}$}, Alberto Morpurgo\textit{$^{a}$}, Enrico Giannini\textit{$^{a}$} and Fabian O. von Rohr$^{\ast}$\textit{$^{a}$}} \\

\includegraphics{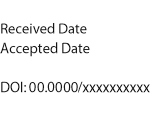} & \noindent\normalsize{In recent years, interest in 2D Janus materials has grown exponentially, particularly with regard to their applications in spintronics and optoelectronic devices. The defining feature of Janus materials is the ordered arrangement of different layer terminations -- creating chemically distinct surfaces and an inherent out-of-plane polarity.
Among the few known Janus materials, \ce{RhSeCl} is particularly intriguing as a rare example of an intrinsic Janus compound. Owing to its exceptional chemical stability, \ce{RhSeCl} offers a promising platform for exploring the physics related to the Janus-structure. However, synthesising large, high-quality crystals of this compound remains a significant challenge. Here, we report a novel synthetic pathway for growing crystals up to \SI{6}{\milli\meter} in lateral size via a two-step self-selecting vapour growth reaction. We further present a comprehensive comparison of newly developed synthesis routes with all previously reported methods for \ce{RhSeCl}. During these investigations, we identified a previously unreported impurity that forms in specific growth pathways and demonstrate how it can be avoided to obtain phase-pure few- and monolayer flakes. We showcase the reproducibility of the process to obtain high-quality, large single-crystals and flakes.} \\

\end{tabular}

 \end{@twocolumnfalse} \vspace{0.6cm}

  ]

\renewcommand*\rmdefault{bch}\normalfont\upshape
\rmfamily
\section*{}
\vspace{-1cm}


\footnotetext{\textit{$^{a}$~Department of Quantum Matter Physics, University of Geneva, 24 Quai Ernest-Ansermet, CH-1211 Geneva, Switzerland.; Tel: +41 22 379 64 78; E-mail: fabian.vonrohr@unige.ch }}


\footnotetext{$^\ddag$ Shared first co-authorship.}

\footnotetext{$^\ast$ Corresponding author.}



\section{Introduction}

Mixed-anion materials, in which two or more different anions coexist within one crystal structure at separate crystallographic sites, have emerged as a versatile platform for designing functional materials with tunable physical properties.\cite{kageyama2018expanding,harada2019heteroanionic} Their chemical flexibility enables fine control over electronic structure, bonding, and symmetry -- leading to phenomena such as ferroelectricity, Rashba splitting, and nonlinear optical effects.\cite{oka2014possible,ishizaka2011giant,zhang2023oxychalcogenides}

Mixed-anion van der Waals (vdW) materials have recently attracted increasing interest, as this particular type of chemistry enables controlled structural symmetry breaking.\cite{gibson2019modular} Representative examples include the magnetic semiconductor \ce{CrSBr},\cite{ziebel2024crsbr,wu2022quasi,lopez2022dynamic}, which displays quasi-one-dimensional transport properties, and \ce{CrPS4}, a layered thiophosphate antiferromagnet with weak and tunable magnetic anisotropy, as well as a pronounced structural and optical anisotropy.\cite{lee2017structural,wang2025configurable} Of particular interest are intrinsic Janus materials, a distinct type of mixed-anion compound in which different anions occupy opposite sides of each layer, breaking the local inversion symmetry of the central atoms.\cite{cheng2013spin,lu2017janus} This separation due to the different determination of the layers leads to an internal electric field, as well as Rashba splitting, and piezoelectricity.\cite{Zhang2020Review} 

The properties of two-dimensional Janus materials were first predicted theoretically and then studied experimentally using a bottom-up approach based on layer engineering. This was realised by a multitude of ways: through asymmetric modification of graphene with halogen atoms or organic molecules with various functional groups; \cite {zhang2013janus, Bissett2014JanusGraphene,li2018recent} by replacing one layer of \ce{S} or \ce{Se} atoms in a monolayer of \ce{MoSe2} or \ce{MoS2}, respectively, to obtain MoSeS;\cite {Zhang2017JanusMoSSe,lu2017janus,liu2025polarization} and through layer-by-layer deposition using epitaxial techniques.\cite{Hajra2020EpitaxialLayers}

The top-down approach, in which bulk crystals can be exfoliated down to a monolayer, is a simple and inexpensive method, but it can only be implemented if large, high-quality crystals of intrinsic Janus materials are grown. Only a few Janus compounds have been discovered so far that crystallise directly with separated ordered anionic layers. Among them are a family of bismuth chalcohalides\cite{sklyadneva2012lattice,Jacimovic2014BiTeCl,akrap2014optical} and breathing kagome Nb$_3Ch$I$_7$ (with $Ch$ = Se, Te).\cite{wang2025longitudinal,yun2025flat} In addition to these materials, a further compound, \ce{RhSeCl}, has recently been found to crystallise in a van der Waals layered structure, in the space group $P6_3mc$, with separated layer determinations of chlorine and selenium anions, i.e., in a Janus structure. This phase was initially discovered in 2018 as an analogue to \ce{RhTeCl},\cite{Kretschmer2019KomplexeRhodiumchalkogenidhalogenide} and was later identified as a non-centrosymmetric 2D Janus material in 2023.\cite{Nowak2023CrystalRhSeCl}

The growth of large \ce{RhSeCl} crystals has proven to be difficult, and the reported syntheses produce only polycrystalline powders or small intergrown crystals. The initial successful approach in 2018 departed from employing \ce{AlCl3} as a chlorine source as used in \ce{RhTeCl}, and instead was completed as a one-step solid-state reaction employing merely \ce{Se} and \ce{RhCl3}, see equation \eqref{eq:1}.\cite{Kretschmer2019KomplexeRhodiumchalkogenidhalogenide} To avoid the formation of \ce{SeCl4} and \ce{Se2Cl2}, \textit{Nowak et al.} adapted this to a chemical-vapour-transport (CVT) synthesis and introduced elemental \ce{Rh} into the reaction, see equation  \eqref{eq:2}. In their CVT, they employed a sink temperature $T_1=\SI{900}{\celsius}$ and source temperature $T_2=\SI{1000}{\celsius}$, as well as a small excess of \ce{RhCl3}.\cite{Nowak2023CrystalRhSeCl} In addition to this conventional CVT approach with a static temperature profile, they performed crystal growth from the gas phase with 27 temperature oscillation cycles.\cite{Nowak2025SyntheseAnaloga} In early 2025, \textit{Liu et al.} adapted the CVT approach by adjusting the \ce{RhCl3} excess to \SI{10}{\percent} to enhance the transport rate of the crystals, and changing the sink temperature to $T_1=\SI{930}{\celsius}$ with maintaining $T_2=\SI{1000}{\celsius}$ as they found that a smaller $\Delta T$ of \SI{70}{\celsius} reduced the thermal stress and the formation of defects in the crystals during growth.\cite{liu2025optimized} All the methods resulted in up to \SI{1}{\milli\meter} laterally-sized crystals.\cite{Nowak2023CrystalRhSeCl,Nowak2025SyntheseAnaloga,liu2025optimized}

\begin{equation}
    \ce{2RhCl3 + 3Se -> 2RhSeCl + SeCl4} \label{eq:1} 
\end{equation}

\begin{equation}
    \ce{2Rh + RhCl3 + 3Se -> 3RhSeCl} \label{eq:2} 
\end{equation}

Here, we present a systematic study of the synthesis and crystal growth of the Janus compound \ce{RhSeCl}. Understanding and exploiting the properties of this intrinsically non-centrosymmetric material requires large, phase-pure, and reproducible single crystals, which in turn can only be obtained through a comprehensive exploration of the growth parameters and optimisation of the processing conditions. We have compared solid-state reactions, chemical vapour transport (CVT), and self-selecting vapour growth (SSVG) and have identified the key parameters that govern crystal quality and size. Building on the proven SSVG method for transition-metal halides,\cite{yan2023self,Lukovkina2024ControllingCo1-xNixI2} we applied this approach to \ce{RhSeCl}, achieving large single crystals. Introducing \ce{SeCl4} as a new starting material further enhanced the chemical purity by eliminating \ce{RhCl3} intergrowths during crystal formation. By combining a two-step optimised SSVG process with the new reagent composition, we have substantially increased the final crystal size. Our results establish practical guidelines for producing high-quality \ce{RhSeCl} crystals and provide a foundation for future physical properties studies and device applications of transition-metal Janus materials.

\section{Experimental Methods} \label{Sec:Experimental}

\subsection{Synthesis}

As starting materials \ce{Rh} powder (Alfa Aesar, \SI{99.9}{\percent}), \ce{RhCl3} powder (ChemPur, \SI{49.9}{\percent} \ce{Rh}), \ce{Se} pieces (Alfa Aesar, \SI{99.999}{\percent}), and \ce{SeCl4} powder (Sigma Aldrich, \SI{35.0}{\percent} \ce{Se}) were used. All chemical handling was carried out inside an argon-filled glovebox due to the air-sensitivity of \ce{SeCl4} and \ce{RhCl3}. All syntheses were performed in quartz ampules with an inner diameter of \SI{10}{\milli\meter} and wall thickness of \SI{1}{\milli\meter}, sealed under dynamic vacuum, with the starting materials totalling a mass of approximately \SI{200}{\milli\gram}. Two main combinations of starting materials were used to determine which would yield the best results. These were: (1) $\ce{Rh}:\ce{RhCl3}:\ce{Se}$ with a molar ratio of $2:1:3$ (with and without \SI{10}{\percent} of \ce{RhCl3} excess) and (2) $\ce{Rh}:\ce{SeCl4}:\ce{Se}$ in a stoichiometric ratio of $4:1:3$. The ampules containing \ce{SeCl4} were initially cooled with ice and then with liquid nitrogen during the sealing of the ampules to prevent the evapouration of \ce{SeCl4} and \ce{Se2Cl2} (details in Results and Discussion section). For crystal growth, two main approaches were applied:

\subsubsection{Chemical Vapour Transport Method}

The \ce{RhSeCl} crystals were first grown by adapting the CVT protocol reported by Novak \textit{et al.}\cite{Nowak2023CrystalRhSeCl} using the two different compositions of the starting materials described above. For all syntheses, \SI{10}{\centi\meter} long quartz ampules were used in a tubular furnace with a temperature difference of $\Delta T=\SI{100}{\celsius}$ with $T_1=\SI{900}{\celsius}$ and $T_2=\SI{1000}{\celsius}$. 

\subsubsection{Self-Selecting Vapour Growth Method}

The SSVG\cite{yan2023self,Lukovkina2024ControllingCo1-xNixI2} approach was adapted for \ce{RhSeCl} using both reagent combinations. Two furnace configurations were tested: a box furnace mounted on its side and a tubular furnace (with one end of the quartz ampule positioned in the middle of the furnace), both providing a very small temperature gradient. For all SSVG syntheses \SI{6}{\centi\meter} long quartz ampules were used. In the box furnace, the following temperature profile was applied: heating from room temperature (RT) with \SI{+50}{\degree\per\hour} up to \SI{1000}{\celsius}, dwelling there for \SI{24}{\hour}, then slowly cooling with \SI{-1}{\degree\per\hour} down to \SI{900}{\celsius}, followed by cooling to RT at \SI{-150}{\degree\per\hour}. For the syntheses in a tubular furnace, apart from the two reagent combination discussed above, we also performed recrystallization of \ce{RhSeCl} (pre-synthesized by SSVG in a ‘flipped’ box furnace) and applied a similar temperature profile but at higher temperatures: heating from RT with \SI{+50}{\degree\per\hour} up to \SI{1100}{\celsius}, dwelling there for \SI{24}{\hour}, then slowly cooling with \SI{-1}{\degree\per\hour} down to \SI{1075}{\celsius}, followed by cooling to RT at \SI{-150}{\degree\per\hour}. In the tubular furnace, a temperature profile without slow cooling was also employed: heating from RT with \SI{+50}{\degree\per\hour} up to \SI{1100}{}, \SI{1090}{}, or \SI{1075}{\celsius}, dwelling at the respective temperatures for \SI{120}{\hour}, followed by cooling to RT with \SI{-150}{\degree\per\hour}. 

Furthermore, alternative SSVG syntheses were carried out using oscillating temperature profiles either between  \SI{1100}{\celsius} and \SI{1000}{\celsius} or between \SI{1000}{\celsius} and \SI{900}{\celsius}, to probe possible crystal growth regions for the two reagent combinations described above. In both cases, the quartz ampules were placed upright in alumina crucibles inside a conventional box furnace.

For the first variation ($1100-\SI{1000}{\celsius}$), the ampule was heated from RT at \SI{+50}{\degree\per\hour} to \SI{1100}{\celsius}, held for \SI{24}{\hour}, and then cooled stepwise to \SI{1075}{}, \SI{1050}{}, \SI{1025}{}, \SI{1000}{}, and finally to \SI{900}{\celsius} at a rate of \SI{-1}{\degree\per\hour}, with each step including a \SI{1}{\hour} dwell at the target temperature. The final cooling to RT was performed at \SI{-150}{\degree\per\hour}.

The second variation ($1000-\SI{900}{\celsius}$) followed the same principle. The ampule was heated from RT at \SI{+50}{\degree\per\hour} to \SI{1000}{\celsius}, held for \SI{24}{\hour}, and then cooled stepwise to \SI{975}{}, \SI{950}{}, \SI{925}{}, and \SI{900}{\celsius}, with \SI{1}{\hour} dwell periods at each step. The sample was then cooled to RT at \SI{-150}{\degree\per\hour}.

The grown crystals were washed under vacuum filtration with acetonitrile to remove any residues of \ce{Se2Cl2}, then water to hydrolyse residues of \ce{SeCl4}, and finally ethanol to dry the crystals. Any individual adaptations to these synthetic approaches, as well as further details and differences, are discussed in the Results and Discussion section.

\subsection{Scanning Electron Microscopy and Elemental Analysis}

The microstructure of the grown crystals was studied with a JEOL JSM-IT800 scanning electron microscope (SEM) equipped with an Oxford Silicon Drift Detector (SDD) X-Max\textsuperscript{N} that was used for the energy-dispersive X-ray spectroscopy (EDS) analysis ($U_{acc}=15-$\SI{20}{\kilo\eV}). Due to the semiconducting nature of \ce{RhSeCl}, thin crystals were used to avoid charging, and larger crystals were cleaved with Kapton tape beforehand. For both the acquisition of the electron images and the collection of the EDS data, the crystals were fixed on a carbon pad or on a silicon wafer for the exfoliated flakes of \ce{RhSeCl}. 

\subsection{Powder X-ray Diffraction}
Powder X-ray diffraction (PXRD) data (Cu-$K\alpha$, $\lambda\approx\SI{154}{\pico\meter}$, $T=\SI{296(3)}{\celsius}$) were collected on a Rigaku SmartLab diffractometer (Rigaku, Debye-Scherrer geometry, D/tex detector). The crystals for the measurements were finely ground with corn starch in an agate mortar to decrease the preferred orientation of the strongly layered material. The powders were then sealed in borosilicate glass capillaries (Hilgenberg GmbH, outer $\varnothing=\SI{0.8}{\milli\meter}$, wall thickness $=\SI{0.01}{\milli\meter}$). Rietveld refinements were performed using TOPAS Academic software.\cite{Coelho2018TOPASAn} 

\subsection{Single Crystal X-ray Diffraction}
Single crystal X-ray diffraction (SCXRD) was measured on a four-circle Oxford Diffraction SuperNova diffractometer (Agilent/Rigaku SE) with mirror optics and an Atlas CCD detector at $T=\SI{100(1)}{\celsius}$ or $T=\SI{250(1)}{\celsius}$. Mo-$K\alpha$ ($\lambda=\SI{71.073}{\pico\meter}$) was used. The crystals were prepared in immersion oil and mounted using Cryoloops. A numerical (Gaussian) absorption correction based on the crystal shape was applied, and the structure was solved with direct methods and subsequent refinements against $F_o^2$ using SHELXL\cite{Sheldrick2015SHELXTDetermination,Sheldrick2015CrystalSHELXL} and Olex2.\cite{Dolomanov2009OLEX2:Program} The integrated CrysAlisPro software was used for data collection and initial treatment of the data.\cite{CrysAlisPRO}

\begin{figure*}[!t]
    \centering
    \includegraphics[width=\textwidth]{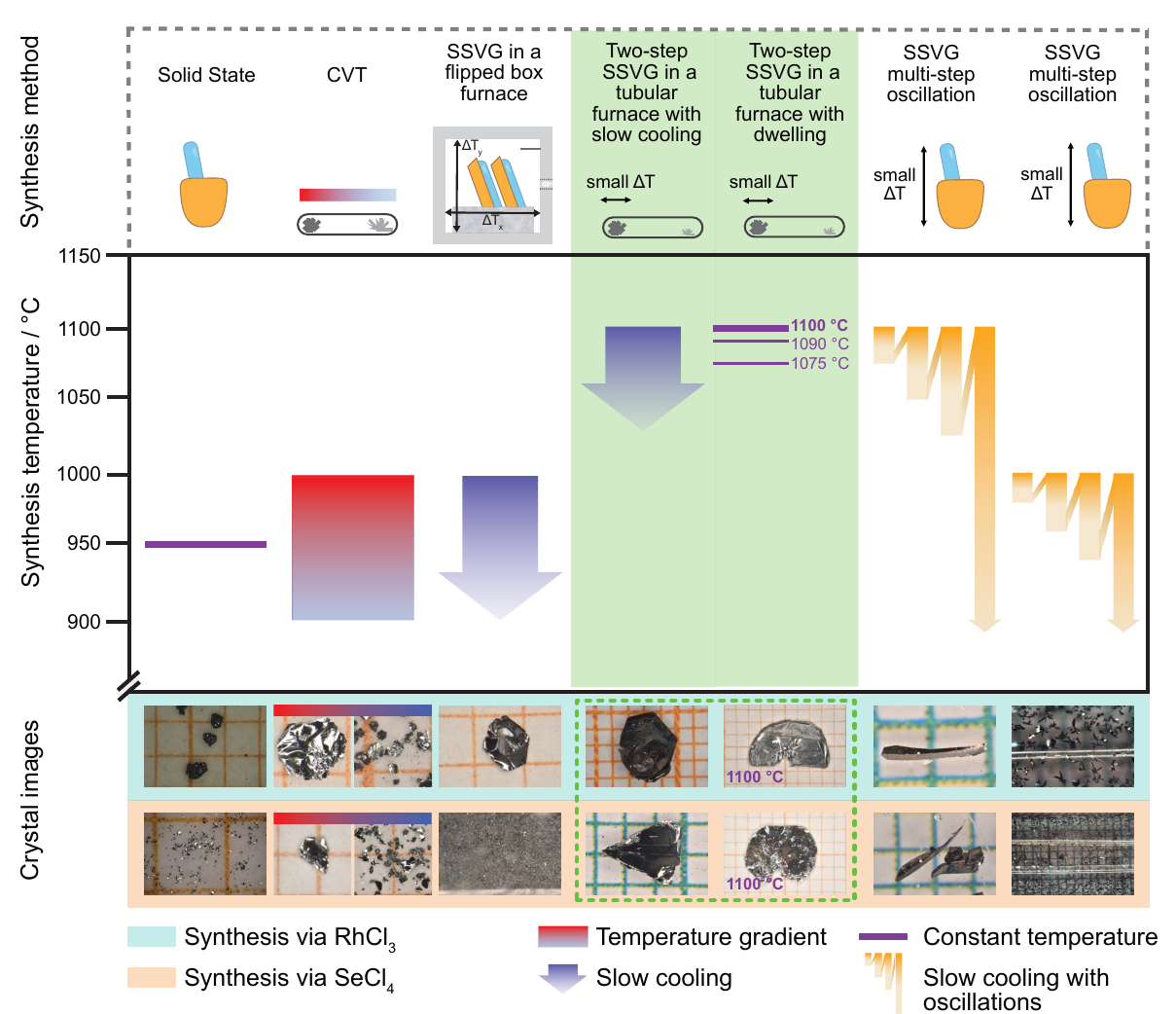}
    \caption{Synthesis map for \ce{RhSeCl} crystal growth. Investigated methods are shown in the upper panel. Temperature profiles for each technique are presented in the middle. Pictures of the \ce{RhSeCl} crystals taken with an optical microscope on millimetre paper at the bottom correspond to the synthesis method above. The turquoise and orange backgrounds represent different synthesis paths: via \ce{Rh}, \ce{Se}, \ce{RhCl3} or \ce{Rh}, \ce{Se}, \ce{SeCl4} reagent combinations, respectively. For the SSVG in a tubular furnace, the shown crystals are obtained by two-step SSVG process starting from \ce{RhCl3} and \ce{SeCl4}, with the first step being SSVG in a ‘flipped’ box furnace. The green background and the dashed frame highlight the best synthesis methods and conditions, which allow the growth of the largest \ce{RhSeCl} crystals.}
    \label{fig:Synthesis map}
\end{figure*}

\subsection{Exfoliation and Raman spectroscopy}

The \ce{RhSeCl} crystals were exfoliated using standard adhesive-tape methods. Thin flakes were transferred to \ce{Si/SiO2} substrates (\SI{285}{\nano\meter} oxide) and examined under an optical microscope to identify suitable regions for Raman measurements. The number of layers was determined by the optical contrast of the flakes relative to the substrate, following established calibration methods. Raman spectra were recorded with a commercial confocal Raman microscope HORIBA LabRAM HR Evolution equipped with a \SI{532}{\nano\meter} continuous-wave laser. A circularly polarised laser beam was focused through an Olympus $\times100$ ($NA=0.9$) objective onto the sample, forming a spot with FWHM $\sim$ \SI{0.3}{\upmu\meter}. To avoid overheating, the laser power at the sample was reduced to \SI{200}{\upmu\watt}, as confirmed by temperature estimates in the laser spot derived from the Stokes/anti-Stokes ratio.

\section{Results and Discussion}

In our exploratory work on the preparation of large \ce{RhSeCl} crystals, we performed a comparative study of different synthesis and growth methods.

In the following, we will compare the reactions that develop from two different combinations of starting materials, one based on \ce{RhCl3} ($\ce{Rh}:\ce{RhCl3}:\ce{Se} = 2:1:3$) and described by equation \ref{eq:2}, another using \ce{SeCl4} ($\ce{Rh}:\ce{SeCl4}:\ce{Se} = 4:1:3$) and described by equation \ref{eq:Rh+Se+SeCl4}.

\begin{equation}
    \ce{4Rh + 3Se + SeCl4 = 4RhSeCl} \label{eq:Rh+Se+SeCl4}
\end{equation}

The use of \ce{SeCl4} as an alternative chlorine source provides a new synthetic pathway for \ce{RhSeCl}, enabling a direct comparison of how precursor chemistry influences crystal growth. We employed both reactions with all processing methods investigated here, which are solid-state reaction, chemical vapour transport (CVT), and self-selecting vapour growth (SSVG). These experiments established a baseline for assessing which approach offers the most reliable route to large, phase-pure single crystals.

\begin{figure*}
    \centering
    \includegraphics[width=\textwidth]{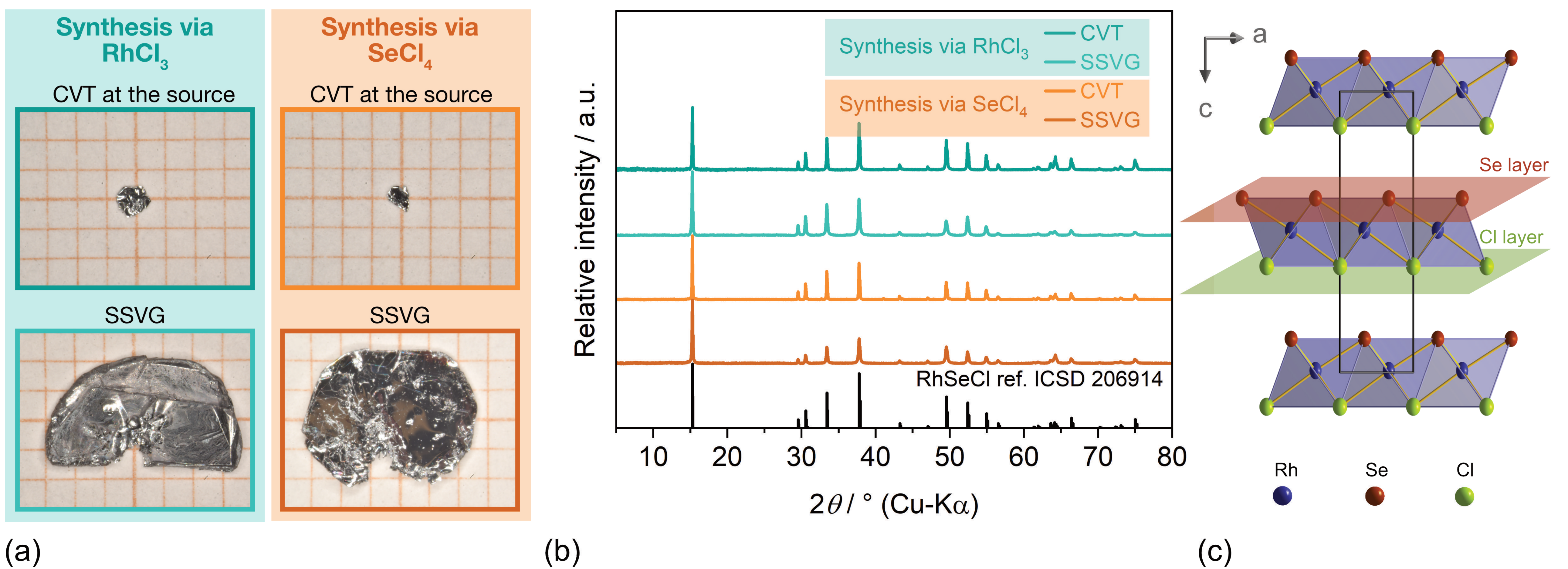}
    \caption{Grown crystals of \ce{RhSeCl} and crystal structure. (a) Comparison of the \ce{RhSeCl} crystals grown by chemical vapour transport (CVT) and self-selecting vapour growth (SSVG). The crystals grown with \ce{RhCl3} are highlighted via a turquoise background, while the crystals grown with \ce{SeCl4} are highlighted via an orange background. (b) PXRD patterns of washed crystals from the different types of reagent combinations and synthesis types. The crystals grown from \ce{RhCl3} are in turquoise, the crystals grown with  \ce{SeCl4} are in orange, and the \ce{RhSeCl} reference (ICSD 206914) is in black. (c) The crystal structure from a refinement of a \ce{RhSeCl} crystal grown via the two-step SSVG synthesis with \ce{RhCl3} as a reagent. The structure here is viewed along the crystallographic \textit{b}-axis to highlight the ordered Janus phase with the \ce{Cl} atoms forming one layer on one side of the central \ce{Rh} and the \ce{Se} atoms forming a layer on the other side.}
    \label{fig:Crystals_PXRD_Structure}
\end{figure*}

In the following, we will refer to the first reagent combination as ‘\ce{RhCl3}’ and the second as ‘\ce{SeCl4}’. It is important to note that when \ce{SeCl4} is used as a starting material, the following reaction occurs at room temperature as soon as \ce{Se} and \ce{SeCl4} are mixed together in a quartz ampule:

\begin{equation}\label{eq:SeCl4}
    \ce{SeCl4 + 3Se -> 2Se2Cl2}
\end{equation}

This could be visually observed through the formation of a brown-red liquid.

\subsection{Chemical Vapour Transport Crystal Growth}

To establish a reference point for crystal quality, we first reproduced the CVT syntheses reported earlier, using, as discussed above, the previously employed \ce{RhCl3}, as well as the here introduced \ce{SeCl4}-based precursor mixtures. In both cases, the reactions yielded high phase purity \ce{RhSeCl} single crystals with comparable morphologies. Interestingly, crystal formation occurred at both ends of the ampules. At the sink ($T_1 = \SI{900}{\celsius}$), we observed intergrown crystal aggregates similar to those previously described,\cite{Nowak2023CrystalRhSeCl} whereas the source ($T_2 = \SI{1000}{\celsius}$) contained larger crystals that reached approximately 1 mm across (Fig.~\ref{fig:Synthesis map}, \ref{fig:Crystals_PXRD_Structure}a). This observation suggests that maintaining a small temperature gradient near the solid source promotes crystal growth, a condition inherently realised in the SSVG method. In contrast, under a large temperature gradient of CVT nucleation was not under control, as a consequence, the crystals could not grow large in size. Encouraged by this and by the earlier success of SSVG in the \ce{Co_{1$-x$}Ni_{$x$}I2}, \ce{RuCl3}, and other transition metal halide systems,\cite{Lukovkina2024ControllingCo1-xNixI2, yan2023self} we next investigated SSVG as an independent growth technique for \ce{RhSeCl}.

\subsection{Self-Selecting Vapour Growth}

To test whether SSVG alone could promote \ce{RhSeCl} crystal growth, we next performed experiments in a ‘flipped’ box furnace, positioned on its side.\cite{yan2023self,Lukovkina2024ControllingCo1-xNixI2} Ampules containing \ce{RhCl3}- or \ce{SeCl4}-based precursor mixtures were heated to \SI{1000}{\celsius}, held for \SI{24}{\hour} and then slowly cooled to \SI{900}{\celsius} at a rate of \SI{-1}{\degree\per\hour} before returning to room temperature.
Indeed, crystals with lateral size up to \SI{1}{\milli\meter} formed when the route with \ce{RhCl3} was used and were comparable in size to those obtained by CVT at the source. For the route with \ce{SeCl4} predominantly phase-pure polycrystalline powder was observed (Fig.~\ref{fig:Synthesis map}). These results demonstrate that SSVG can be utilised to synthesise \ce{RhSeCl} single crystals.

\subsection{Crystal Growth Optimization}

Since we succeeded in growing small \ce{RhSeCl} crystals via SSVG technique in the ‘flipped’ furnace, we then explored whether higher temperatures and a different temperature gradient could further enhance growth. To achieve this, we carried out SSVG in short ampules (\SI{6}{\centi\meter}), with the end containing the precursors placed in the centre of a one-zone tubular furnace. This provided a small thermal gradient without inducing undesired CVT transport. We tested three precursor systems: \ce{RhCl3}-, \ce{SeCl4}-based mixtures, and polycrystalline \ce{RhSeCl} pre-synthesised via SSVG in a ‘flipped’ furnace (without opening the ampule between the syntheses). These experiments were carried out in a similar temperature profile as for the previously outlined SSVG setup, but the temperatures were increased to $1100-\SI{1075}{\celsius}$. Under these conditions, all reactions produced larger crystals ($2-\SI{4}{\milli\meter}$), and the best results were obtained with pre-synthesised via SSVG in the ‘flipped’ furnace \ce{RhSeCl} as the starting material (Fig.~\ref{fig:Synthesis map}). These findings showed that higher growth temperatures promote crystal enlargement, establishing a two-step SSVG route -- initial polycrystalline formation followed by regrowth in a tubular furnace -- as the most effective strategy for producing large \ce{RhSeCl} crystals.

Building on the observation that higher growth temperatures improved crystal size, we next optimised the two-step SSVG synthesis by introducing an isothermal dwelling step in the tubular furnace. Using pre-synthesised in a ‘flipped’ furnace  \ce{RhSeCl} as starting material, we examined the dwell temperatures of \SI{1075}{}, \SI{1090}{}, and \SI{1100}{\celsius} with a duration of \SI{120}{\hour}. The \SI{1100}{\celsius} annealing proved to be the most effective, yielding the largest crystals up to $0.4\times5\times$\SI{6}{\milli\meter} (Fig.~\ref{fig:Synthesis map}, \ref{fig:Crystals_PXRD_Structure}), while even at lower temperatures, crystals exceeding \SI{2}{\milli\meter} were readily obtained. Maintaining a closed ampule between the two synthesis steps was essential: reopening or washing the intermediate product led to incongruent melting and decomposition into \ce{RhSe}, \ce{Rh}, and \ce{SeCl4} (see ESI \dag). This is likely because sealed tubes retain volatile \ce{Cl}- and \ce{Se}-bearing species that act as transport agents during crystal growth. Attempts at higher dwell temperatures (\SI{>1100}{\celsius}) again produced incongruent melts. Large crystals grown under the best conditions often exhibited a small cavity at the base, suggesting growth from a partially molten phase. Altogether, these experiments identify \SI{1100}{\celsius} as the optimal temperature for reproducible growth of large \ce{RhSeCl} crystals via the two-step SSVG approach.

\subsection{Self-Selecting Vapour Growth for Crystals with Needle-like Morphology}

As a final set of experiments, we decided to see whether SSVG reactions where the temperature was oscillated in the previously determined crystal growth regions would lead to any significant improvement or change in the crystal growth. In this case, we carried out the experiments as outlined in the Experimental section above, oscillating the temperature either between 1100 and \SI{1000}{\celsius} or between 1000 and \SI{900}{\celsius}. In both cases, the main difference in comparison to all the other synthetic approaches investigated here lay in the morphology of the grown crystals. While the size of individual crystals was comparable to previously reported results, the crystals grew preferentially in one direction, taking on a needle-like appearance, as can be seen in the last two columns of Fig.~\ref{fig:Synthesis map}. Interestingly, crystal growth was also not localised at one end of the ampule, but was rather consistent across the walls, indicating that there was constant movement and deposition of \ce{RhSeCl} crystal nuclei through the temperature oscillations, demonstrating an SSVG type of growth in a natural vertical temperature gradient of a box furnace. Despite their different morphology, these crystals were found to be structurally identical to the other \ce{RhSeCl} crystals obtained, as proven by SCXRD measurements (Tables S1 and S6, ESI\dag).

\subsection{Comparison of the Crystal Growth Methods}

Comparing all crystal-growth approaches reveals how both the reaction chemistry and the thermal environment govern the formation of high-quality \ce{RhSeCl} crystals. Conventional CVT methods, as previously reported in the literature, reliably produces \ce{RhSeCl} crystals, however, their size remains limited. CVT growth for RhSeCl occurs primarily through vapour transport between the hot and cold ends of the ampule, leading to agglomerates of small crystallites, while additional evaporation-condensation processes at the source can occasionally yield larger, isolated crystals. In contrast, the SSVG technique in a ‘flipped’ box furnace enables growth directly within a minimal temperature gradient, and when applied using \ce{RhCl3}-based precursors, it yields crystals comparable in size to those obtained in CVT reactions at the source. 

Increasing the temperature and changing the temperature gradient within short ampules in a tubular furnace markedly improved the crystal size, indicating that crystal growth in this system is favoured by higher vapour pressures and a partially molten phase. The most significant improvement in \ce{RhSeCl} growth was achieved in the two-step SSVG process: first generating polycrystalline \ce{RhSeCl} via SSVG in a ‘flipped’ furnace and then regrowing crystals at elevated temperature in a tubular furnace. Using this two-step process, we could grow RhSeCl crystals up to \SI{6}{\milli\meter} in lateral size for the synthesis with a dwelling step at \SI{1100}{\celsius}. Maintaining a sealed environment between steps was essential: opening the ampules disrupted the internal vapour balance and led to incongruent melting.

Taken together, these results demonstrate that successful \ce{RhSeCl} crystal growth depends on ensuring an optimal combination of high temperature, limited temperature gradient, and closed-system vapour chemistry. The two-step SSVG approach thus combines the advantages of solid-state synthesis and vapour transport, providing a reproducible pathway to large, phase-pure \ce{RhSeCl} single crystals suitable for further physical characterisation and device fabrication. The overall trends and optimal conditions are summarised in the synthesis map (Fig.~\ref{fig:Synthesis map}), with alternative routes detailed in the ESI\dag.

\subsection{Crystal structure}

Powder X-ray diffraction of all our washed samples predominantly showcased the presence of the phase-pure \ce{RhSeCl} Janus phase. Rietveld refinements of the diffraction patterns obtained for each synthetic approach confirmed an excellent fit to the previously reported structure model of \ce{RhSeCl}, i.e., for the samples grown with \ce{SeCl4} from our two-step SSVG approach we obtained cell parameters of $a=\SI{348.4(2)}{\pico\meter}$ and $c=\SI{1156.9(7)}{\pico\meter}$ in the space group $P6_3mc$. PXRD patterns for the CVT-grown crystals and for the largest \ce{RhSeCl} crystals produced in this work are shown in Fig.~\ref{fig:Crystals_PXRD_Structure}, while the data for all other syntheses, together with the Rietveld analyses, are presented in the ESI\dag.

\begin{figure*}[!t]
    \centering
    \includegraphics[width=\textwidth]{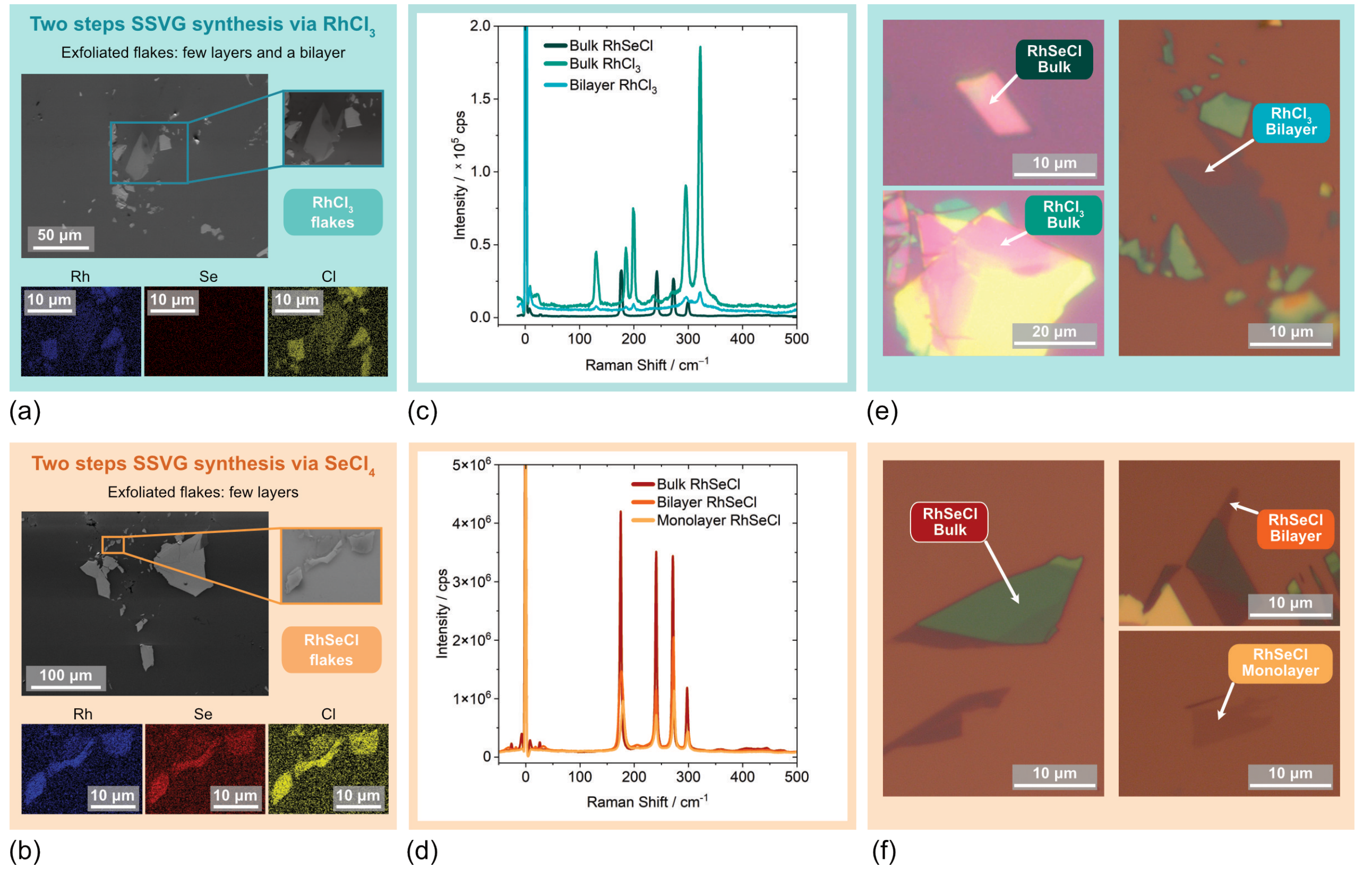}
    \caption{Elemental analysis and Raman spectra of bulk and exfoliated \ce{RhSeCl} crystals grown via two-step SSVG starting from two different reagent combinations. The upper panels show results for \ce{RhSeCl} crystals grown from \ce{RhCl3}, the bottom ones from \ce{SeCl4}. (a,b) Micrograph and elemental analysis of exfoliated \ce{RhSeCl} crystals. The dominant phase observed is \ce{RhCl3} for crystals grown from \ce{RhCl3}, while for crystals grown from \ce{SeCl4} it is \ce{RhSeCl}. (c,d) Comparison of Raman spectra of bulk and exfoliated \ce{RhSeCl} crystals. (e,f) Polarised optical microscope pictures of \ce{RhSeCl} flakes on which Raman measurements were performed.}
    \label{fig:SEM_EDX_Raman}
\end{figure*}

We tested the single-crystals of the individual methods by SCXRD to verify the crystal quality, and we found all of them to crystallise in the hexagonal space group $P6_3mc$. The crystals grown by the herein newly reported SSVG approach with \ce{SeCl4} as a precursor have lattice parameters of $a=\SI{349.11(1)}{\pico\meter}$ and $c=\SI{1158.89(3)}{\pico\meter}$, while the single-crystals grown from \ce{SeCl4} that take on a needle-like morphology (see Section 3.4) show lattice parameters of $a=\SI{349.15(3)}{\pico\meter}$ and $c=\SI{1158.29(12)}{\pico\meter}$. In both cases, these parameters vary merely by approximately \SI{0.1}{\percent} from those found in earlier work and hence no significant differences in the crystal structures could be observed (see Table S2, ESI\dag).

As expected for these Janus structures, as seen in a representative crystal that was grown from \ce{SeCl4} with the two-step SSVG method, the central \ce{Rh} atom is coordinated by three \ce{Cl} atoms with a distance of \SI{250.84(23)}{\pico\meter} and by three \ce{Se} atoms with a distance of \SI{240.34(8)}{\pico\meter} in the expected facial arrangement with the anions ordering according to their type on opposite sides (see Fig.~\ref{fig:Crystals_PXRD_Structure}). These bond lengths only vary minimally from the previously recorded crystal structure of \ce{RhSeCl} ($<\SI{0.1}{\percent}$). Furthermore, the residual electron density peaks fall within the range of $\pm1$ to 2 electrons~$\times10^{-6}$ \unit{pm^{-3}}, mostly within $0.5-\SI{2}{\angstrom}$ of \ce{Rh}, or $0.5-\SI{1.5}{\angstrom}$ of \ce{Se}. The higher residual electron density peaks near the metal cation (\ce{Rh}) are expected, especially considering the size of \ce{Rh}. 
Although the peaks are close to \ce{Rh} and \ce{Se}, they do not indicate disorder, as they are below $5$ electrons~$\times10^{-6}$ \unit{pm^{-3}} and no negative density is observed on the atomic sites. Additionally, the Debye-Waller factors remain reasonable at both 250 and \SI{100}{\kelvin} (see Table S2, ESI\dag), and also do not indicate any substitutional or dynamic disorder.

We also investigated the possibility of any substitutional disorder by refining mixed \ce{Se} and \ce{Cl} sites on both atomic positions. However, this resulted in no significant change in the occupancy of the 'original' atom on that position: in the case of the \ce{Cl} position, i.e., the $2a$ Wyckhoff position, the occupancy remained $99.5$ to $\SI{99.8}{\percent}$ \ce{Cl} and on the \ce{Se} position, i.e., the $2b$ Wyckhoff position, the occupancy remained approx. $\SI{99.5}{\percent}$ \ce{Se}, and in both cases, there was no significant change in the refinement parameters. All of this further confirms that there is very little to no intermixing of the \ce{Se} and \ce{Cl} sites, and that the Janus structure remains. 

And finally, we also checked whether any twinning could be observed. First, the possibility of non-merohedral twinning was examined, however, here we found that at most $5 \unit{\percent}$ of the reflections could be assigned to a further twin component. Due to the small presence of the twinned component and the significant negative impact on the structure solutions where a twin component was present, we did not include it in the final structure solution, especially as other warning signs of non-merohedral twinning, i.e., the $F_o$ values of the most disagreeable reflections are not significantly larger than the $F_c$ values, nor is the $K=mean(F_o^2)/mean(F_c^2)$ systematically high for the low-intensity reflections, were not present. 

And finally, the possibility of twinning by merohedry was also investigated, e.g., in the crystal grown via the CVT method from \ce{SeCl4} had a potential merohedric twin with a BASF value of $0.02(4)$ and a twin law of $(-1~0~0~|~1~1~0~|~0~0~-1)$. However, when this is taken into account in the refinement of the structure solution, it has next to no effect on the refinement parameters, which also indicates that there is no significant twinning by merohedry. All of this together indicates that despite small, individual crystal-dependent factors, all of the synthetic methods that we have outlined here produce high-quality single crystals. 

\subsection{Exfoliation of \ce{RhSeCl}}

One of the initial aims of producing larger crystals was to investigate their viability in device applications. As reported in both previous works \cite{Nowak2023CrystalRhSeCl,liu2025optimized} and confirmed here by our CVT-grown crystals, conventional vapour transport only provides small crystals at the lower limit of what is required in terms of exfoliation of thin flakes. Following the successful growth of larger crystals via the two-step SSVG method, we began investigations to determine whether these crystals of \ce{RhSeCl} were also suitable for potential device fabrication. We could successfully exfoliate the \ce{RhSeCl} van der Waals material using the standard mechanical scotch-tape method. We reproducibly obtained large flakes down to few-layers and monolayer thicknesses, which indicates that the crystals grown via the two-step SSVG route are indeed suitable for Janus-type device fabrications. To assess the structural and chemical uniformity of these exfoliated layers, we examined both bulk and few-layer flakes using EDX and Raman spectroscopy.

For crystals with \ce{RhCl3} as the starting reagent in the two-step SSVG synthesis, we observed two different phases with Raman spectroscopy  (see Fig.~\ref{fig:SEM_EDX_Raman}) after being exfoliated. We found that these \ce{RhSeCl} crystals in fact contain some intergrown layers of \ce{RhCl3} which only became visible once we began to exfoliate the crystals down to few-layers (bulk crystals were checked by PXRD, SCXRD and EDX prior to exfoliation, see ESI\dag). We confirmed this through EDS analysis of the same bilayers and bulk flakes, as can be seen in Fig.~\ref{fig:SEM_EDX_Raman} (a).

This impurity becomes visible after exfoliation, probably because \ce{RhCl3} exfoliates more easily than \ce{RhSeCl}, and due to this, even tiny amounts of intergrown \ce{RhCl3} result in the dominant appearance of \ce{RhCl3} flakes.

When we repeated the same exfoliation experiments with the crystals grown starting from \ce{SeCl4}, we observed that they did not contain any inclusions of \ce{RhCl3}, as can be seen in Figs.~\ref{fig:SEM_EDX_Raman}(b), (d), and (f). Even down to the monolayer, only pure \ce{RhSeCl} was found, which was reproducible between all batches grown starting from \ce{SeCl4} via the two-step SSVG synthesis route. Hence, the crystals grown as such are the only ones that are both large enough to be used in device applications and for exfoliation experiments, and in addition to this, they are free of inherently intergrown impurities that would hinder their suitability.

As a result of both the optimizations in crystal synthesis with regard to the method and the combination of reagents, together with the initial EDX and Raman investigations of the exfoliated bilayer and few-layer flakes, we can say that the crystals grown from \ce{SeCl4} by the two-step SSVG method yield the best, reproducible and most promising samples for further explorations towards device fabrication. 

\section{Conclusions}

We have thoroughly investigated the crystal growth process of the recently discovered and highly promising Janus material \ce{RhSeCl}. We grew single crystals using various methods and two different reagent combinations with \ce{RhCl3} and \ce{SeCl4} as the chlorine source.
First, we could confirm and reproduce the results previously obtained by the CVT method and proved that the crystal size remains very limited while using this technique. We found that the two-step SSVG process (pre-synthesis performed in a ‘flipped’ furnace, followed by recrystallisation at higher temperatures in a tubular furnace) provides reproducible crystals up to $0.4\times5\times$\SI{6}{\milli\meter^3}. 
Moreover, the choice of the precursor phases was also found to be critical. We observed the intergrowth of thin \ce{RhCl3} flakes when this compound is used as a reagent. The volume fraction of these \ce{RhCl3} flakes was not visible by bulk-sensitive probes, but was sufficient to dramatically affect the device fabrication. We propose an alternative synthesis approach in which \ce{SeCl4} is used as a precursor, and we proved that in this case, no intergrown impurities were detected and the exfoliated flakes all show a high and reproducible quality. Taken together, we find the optimal crystal growth conditions to be a two-step SSVG process, starting with the synthesis of polycrystalline \ce{RhSeCl} from \ce{Rh}, \ce{Se}, and \ce{SeCl4} in a ‘flipped’ furnace via slow cooling from \SI{1000}{\celsius} to \SI{900}{\celsius}, followed by recrystallisation by dwelling at \SI{1100}{\celsius} (or slow cooling from \SI{1100}{\celsius} to \SI{1075}{\celsius}) in a tubular furnace to yield large single crystals. 
These optimised conditions now provide reliable access to large, structurally pure \ce{RhSeCl} crystals, creating a solid foundation for meaningful device-level studies.


\section*{Conflicts of interest}

There are no conflicts to declare.

\section*{Acknowledgements}

This work was supported by the Swiss National Science Foundation under Grants No. PCEFP2\_194183 and No. 200021-204065.



\balance


\bibliography{RhSeCl} 

@article{kageyama2018expanding,
  title={Expanding frontiers in materials chemistry and physics with multiple anions},
  author={Kageyama, Hiroshi and Hayashi, Katsuro and Maeda, Kazuhiko and Attfield, J Paul and Hiroi, Zenji and Rondinelli, James M and Poeppelmeier, Kenneth R},
  journal={Nature communications},
  volume={9},
  number={1},
  pages={772},
  year={2018},
  publisher={Nature Publishing Group UK London}
}

@article{lee2017structural,
  title={Structural and optical properties of single-and few-layer magnetic semiconductor \ce{CrPS4}},
  author={Lee, Jinhwan and Ko, Taeg Yeoung and Kim, Jung Hwa and Bark, Hunyoung and Kang, Byunggil and Jung, Soon-Gil and Park, Tuson and Lee, Zonghoon and Ryu, Sunmin and Lee, Changgu},
  journal={ACS nano},
  volume={11},
  number={11},
  pages={10935--10944},
  year={2017},
  publisher={ACS Publications}
}

@article{wu2022quasi,
  title={Quasi-1D electronic transport in a 2D magnetic semiconductor},
  author={Wu, Fan and Guti{\'e}rrez-Lezama, Ignacio and L{\'o}pez-Paz, Sara A and Gibertini, Marco and Watanabe, Kenji and Taniguchi, Takashi and von Rohr, Fabian O and Ubrig, Nicolas and Morpurgo, Alberto F},
  journal={Advanced Materials},
  volume={34},
  number={16},
  pages={2109759},
  year={2022},
  publisher={Wiley Online Library}
}

@article{gibson2019modular,
  title={Modular design via multiple anion chemistry of the high mobility van der Waals semiconductor \ce{Bi4O4SeCl2}},
  author={Gibson, Quinn D and Manning, Troy D and Zanella, Marco and Zhao, Tianqi and Murgatroyd, Philip AE and Robertson, Craig M and Jones, Leanne AH and McBride, Fiona and Raval, Rasmita and Cora, Furio and others},
  journal={Journal of the American Chemical Society},
  volume={142},
  number={2},
  pages={847--856},
  year={2019},
  publisher={ACS Publications}
}

@article{ishizaka2011giant,
  title={Giant Rashba-type spin splitting in bulk BiTeI},
  author={Ishizaka, K and Bahramy, MS and Murakawa, H and Sakano, M and Shimojima, T and Sonobe, T and Koizumi, K and Shin, S and Miyahara, H and Kimura, A and others},
  journal={Nature materials},
  volume={10},
  number={7},
  pages={521--526},
  year={2011},
  publisher={Nature Publishing Group UK London}
}

@article{zhang2023oxychalcogenides,
  title={Oxychalcogenides: A Promising Class of Materials for Nonlinear Optical Crystals with Mixed-Anion Groups},
  author={Zhang, Yujie and Wu, Hongping and Hu, Zhanggui and Yu, Hongwei},
  journal={Chemistry--A European Journal},
  volume={29},
  number={17},
  pages={e202203597},
  year={2023},
  publisher={Wiley Online Library}
}

@article{oka2014possible,
  title={Possible ferroelectricity in perovskite oxynitride \ce{SrTaO2N} epitaxial thin films},
  author={Oka, Daichi and Hirose, Yasushi and Kamisaka, Hideyuki and Fukumura, Tomoteru and Sasa, Kimikazu and Ishii, Satoshi and Matsuzaki, Hiroyuki and Sato, Yukio and Ikuhara, Yuichi and Hasegawa, Tetsuya},
  journal={Scientific reports},
  volume={4},
  number={1},
  pages={4987},
  year={2014},
  publisher={Nature Publishing Group UK London}
}

@article{harada2019heteroanionic,
  title={Heteroanionic materials by design: Progress toward targeted properties},
  author={Harada, Jaye K and Charles, Nenian and Poeppelmeier, Kenneth R and Rondinelli, James M},
  journal={Advanced Materials},
  volume={31},
  number={19},
  pages={1805295},
  year={2019},
  publisher={Wiley Online Library}
}

@article{Coelho2018TOPASAn,
    title = {{TOPAS and TOPAS-Academic: An optimization program integrating computer algebra and crystallographic objects written in C++: An}},
    year = {2018},
    journal = {Journal of Applied Crystallography},
    author = {Coelho, Alan A.},
    number = {1},
    month = {2},
    pages = {210--218},
    volume = {51},
    publisher = {Wiley-Blackwell},
    doi = {10.1107/S1600576718000183},
    issn = {16005767}
}

@article{Lukovkina2024ControllingCo1-xNixI2,
    title = {{Controlling the Magnetic Properties of the van der Waals Multiferroic Crystals Co1-xNixI2}},
    year = {2024},
    journal = {Chemistry of Materials},
    author = {Lukovkina, Anastasiia and L{\'{o}}pez-Paz, Sara A. and Besnard, Céline and Guenee, Laure and von Rohr, Fabian O. and Giannini, Enrico},
    number = {12},
    month = {6},
    pages = {6237--6245},
    volume = {36},
    publisher = {American Chemical Society},
    doi = {10.1021/acs.chemmater.4c01053},
    issn = {15205002},
    arxivId = {2406.09146}
}

@article{Nowak2023CrystalRhSeCl,
    title = {{Crystal growth of the 2D Janus rhodium chalcohalide RhSeCl}},
    year = {2023},
    journal = {Inorganic Chemistry Frontiers},
    author = {Nowak, Domenic and Valldor, Martin and Rubrecht, Bastian and Froeschke, Samuel and Eltoukhy, Samar and B{\"{u}}chner, Bernd and Hampel, Silke and Gr{\"{a}}{\ss}ler, Nico},
    number = {10},
    month = {2},
    pages = {2911--2916},
    volume = {10},
    publisher = {Royal Society of Chemistry},
    doi = {10.1039/d2qi02699f},
    issn = {20521553}
}

@article{Sheldrick2015CrystalSHELXL,
    title = {{Crystal structure refinement with SHELXL}},
    year = {2015},
    journal = {Acta Crystallographica Section C: Structural Chemistry},
    author = {Sheldrick, George M.},
    month = {1},
    pages = {3--8},
    volume = {71},
    publisher = {International Union of Crystallography},
    doi = {10.1107/S2053229614024218},
    issn = {20532296},
    pmid = {25567568}
}

@article{ziebel2024crsbr,
  title={CrSBr: an air-stable, two-dimensional magnetic semiconductor},
  author={Ziebel, Michael E and Feuer, Margalit L and Cox, Jordan and Zhu, Xiaoyang and Dean, Cory R and Roy, Xavier},
  journal={Nano letters},
  volume={24},
  number={15},
  pages={4319--4329},
  year={2024},
  publisher={ACS Publications}
}

@article{li2018recent,
  title={Recent progress of Janus 2D transition metal chalcogenides: from theory to experiments},
  author={Li, Ruiping and Cheng, Yingchun and Huang, Wei},
  journal={Small},
  volume={14},
  number={45},
  pages={1802091},
  year={2018},
  publisher={Wiley Online Library}
}

@article{liu2025polarization,
  title={Polarization-Field-Induced Inequivalent Exciton Dynamics in Janus MoSeS/\ce{MoSe2} Heterostructures},
  author={Liu, Mengyu and Wu, Wei and Chen, Zilong and Zhang, Yuxiang and Yu, Xingcheng and Yang, Shunhang and Wang, Hao and Xu, Feiya and Chen, Li and Li, Xu and others},
  journal={Nano Letters},
  volume={25},
  number={14},
  pages={5723--5730},
  year={2025},
  publisher={ACS Publications}
}

@article{wang2025configurable,
  title={Configurable antiferromagnetic domains and lateral exchange bias in atomically thin \ce{CrPS4}},
  author={Wang, Yu-Xuan and Graham, Thomas KM and Rama-Eiroa, Ricardo and Islam, Md Ariful and Badarneh, Mohammad H and Nunes Gontijo, Rafael and Tiwari, Ganesh Prasad and Adhikari, Tibendra and Zhang, Xin-Yue and Watanabe, Kenji and others},
  journal={Nature Materials},
  pages={1--10},
  year={2025},
  publisher={Nature Publishing Group UK London}
}

@article{lopez2022dynamic,
  title={Dynamic magnetic crossover at the origin of the hidden-order in van der Waals antiferromagnet CrSBr},
  author={L{\'o}pez-Paz, Sara A and Guguchia, Zurab and Pomjakushin, Vladimir Y and Witteveen, Catherine and Cervellino, Antonio and Luetkens, Hubertus and Casati, Nicola and Morpurgo, Alberto F and von Rohr, Fabian O},
  journal={Nature Communications},
  volume={13},
  number={1},
  pages={4745},
  year={2022},
  publisher={Nature Publishing Group UK London}
}

@article{Hajra2020EpitaxialLayers,
    title = {{Epitaxial synthesis of highly oriented 2D Janus rashba semiconductor BiTeCl and BiTeBr layers}},
    year = {2020},
    journal = {ACS Nano},
    author = {Hajra, Debarati and Sailus, Renee and Blei, Mark and Yumigeta, Kentaro and Shen, Yuxia and Tongay, Sefaattin},
    number = {11},
    month = {11},
    pages = {15626--15632},
    volume = {14},
    publisher = {American Chemical Society},
    doi = {10.1021/acsnano.0c06434},
    issn = {1936086X},
    pmid = {33090763}
}

@article{lu2017janus,
  title={Janus monolayers of transition metal dichalcogenides},
  author={Lu, Ang-Yu and Zhu, Hanyu and Xiao, Jun and Chuu, Chih-Piao and Han, Yimo and Chiu, Ming-Hui and Cheng, Chia-Chin and Yang, Chih-Wen and Wei, Kung-Hwa and Yang, Yiming and others},
  journal={Nature nanotechnology},
  volume={12},
  number={8},
  pages={744--749},
  year={2017},
  publisher={Nature Publishing Group UK London}
}

@article{cheng2013spin,
  title={Spin-orbit--induced spin splittings in polar transition metal dichalcogenide monolayers},
  author={Cheng, YC and Zhu, ZY and Tahir, Muhammad and Schwingenschl{\"o}gl, Udo},
  journal={Europhysics Letters},
  volume={102},
  number={5},
  pages={57001},
  year={2013},
  publisher={IOP Publishing}
}

@phdthesis{Kretschmer2019KomplexeRhodiumchalkogenidhalogenide,
    title = {{Komplexe Halogenidorhodate und Rhodiumchalkogenidhalogenide}},
    year = {2019},
    author = {Kretschmer, Jan},
    school = {Rheinische Friedrich-Wilhelms-Universit{\"{a}}t Bonn},
    language = {German}
}

@article{Dolomanov2009OLEX2:Program,
    title = {{OLEX2: A complete structure solution, refinement and analysis program}},
    year = {2009},
    journal = {Journal of Applied Crystallography},
    author = {Dolomanov, Oleg V. and Bourhis, Luc J. and Gildea, Richard J. and Howard, Judith A.K. and Puschmann, Horst},
    number = {2},
    pages = {339--341},
    volume = {42},
    doi = {10.1107/S0021889808042726},
    issn = {00218898}
}

@article{liu2025optimized,
  title={Optimized Synthesis and Characterization of Janus RhSeCl with Uniform Anionic Valences, Nonlinear Optical and Optoelectronic Properties},
  author={Liu, Kefeng and Sun, Xuelian and Cheng, Puxin and Li, Zhiteng and Li, Penghui and Jia, Donghan and Zhao, Shijing and Yang, Xin and Wang, Xinyu and Ye, Liangting and others},
  journal={Advanced Science},
  pages={e05279},
  year={2025},
  volume = {12},
  number = {34}
}

@article{yan2023self,
  title={Self-selecting vapor growth of transition-metal-halide single crystals},
  author={Yan, J-Q and McGuire, Michael A},
  journal={Physical Review Materials},
  volume={7},
  number={1},
  pages={013401},
  year={2023},
  publisher={APS}
}

@article{Sheldrick2015SHELXTDetermination,
    title = {{SHELXT - Integrated space-group and crystal-structure determination}},
    year = {2015},
    journal = {Acta Crystallographica Section A: Foundations of Crystallography},
    author = {Sheldrick, George M.},
    number = {1},
    month = {1},
    pages = {3--8},
    volume = {71},
    publisher = {International Union of Crystallography},
    doi = {10.1107/S2053273314026370},
    issn = {16005724}
}

@phdthesis{Nowak2025SyntheseAnaloga,
    title = {{Synthese, Charakterisierung und Anionenordnung des 2D Janusmaterials RhSeCl und seiner strukturellen Analoga}},
    year = {2025},
    author = {Nowak, Domenic},
    publisher = {Technische Universit{\"{a}}t Dresden},
    school = {Technische Universit{\"{a}}t Dresden},
    address = {Dresden},
    language = {German}
}

@article{Bissett2014JanusGraphene,
   author = {Mark A. Bissett and Yuichiro Takesaki and Masaharu Tsuji and Hiroki Ago},
   doi = {10.1039/c4ra09724f},
   issn = {20462069},
   issue = {94},
   journal = {RSC Advances},
   month = {10},
   pages = {52215-52219},
   publisher = {Royal Society of Chemistry},
   title = {Increased chemical reactivity achieved by asymmetrical 'Janus' functionalisation of graphene},
   volume = {4},
   year = {2014}
}

@article{zhang2013janus,
  title={Janus graphene from asymmetric two-dimensional chemistry},
  author={Zhang, Liming and Yu, Jingwen and Yang, Mingmei and Xie, Qin and Peng, Hailin and Liu, Zhongfan},
  journal={Nature Communications},
  volume={4},
  number={1},
  pages={1443},
  year={2013},
  publisher={Nature Publishing Group UK London}
}

@article{Zhang2017JanusMoSSe,
   author = {Jing Zhang and Shuai Jia and Iskandar Kholmanov and Liang Dong and Dequan Er and Weibing Chen and Hua Guo and Zehua Jin and Vivek B. Shenoy and Li Shi and Jun Lou},
   doi = {10.1021/acsnano.7b03186},
   issn = {1936086X},
   issue = {8},
   journal = {ACS Nano},
   month = {8},
   pages = {8192-8198},
   pmid = {28771310},
   publisher = {American Chemical Society},
   title = {Janus Monolayer Transition-Metal Dichalcogenides},
   volume = {11},
   year = {2017}
}

@article{sklyadneva2012lattice,
  title={Lattice dynamics of bismuth tellurohalides},
  author={Sklyadneva, I Yu and Heid, R and Bohnen, K-P and Chis, V and Volodin, VA and Kokh, KA and Tereshchenko, OE and Echenique, PM and Chulkov, EV},
  journal={Physical Review B—Condensed Matter and Materials Physics},
  volume={86},
  number={9},
  pages={094302},
  year={2012},
  publisher={APS}
}

@article{akrap2014optical,
  title={Optical properties of BiTeBr and BiTeCl},
  author={Akrap, Ana and Teyssier, J{\'e}r{\'e}mie and Magrez, Arnaud and Bugnon, Philippe and Berger, Helmuth and Kuzmenko, Alexey B and Van Der Marel, Dirk},
  journal={Physical Review B},
  volume={90},
  number={3},
  pages={035201},
  year={2014},
  publisher={APS}
}

@article{Jacimovic2014BiTeCl,
   author = {J. Jacimovic and X. Mettan and A. Pisoni and R. Gaal and S. Katrych and L. Demko and A. Akrap and L. Forro and H. Berger and P. Bugnon and A. Magrez},
   doi = {10.1016/j.scriptamat.2013.12.017},
   issn = {13596462},
   journal = {Scripta Materialia},
   month = {4},
   pages = {69-72},
   title = {Enhanced low-temperature thermoelectrical properties of BiTeCl grown by topotactic method},
   volume = {76},
   year = {2014}
}

@article{yun2025flat,
  title={Flat-Band Electronic Bipolarity in a Janus and Kagome van der Waals Semiconductor Nb3TeI7},
  author={Yun, Jo Hyun and Sung, Minki and Choi, Minhyuk and Kim, Kyoo and Yang, Wooin and Kim, Dowook and Kim, Min Joong and Her, Sung-Hyuk and Choi, Si-Young and Kim, Tae-Hwan and others},
  journal={Advanced Materials},
  volume={37},
  number={9},
  pages={2415045},
  year={2025},
  publisher={Wiley Online Library}
}

@article{wang2025longitudinal,
  title={Longitudinal Piezoelectricity and Polarization-Insensitive Oxidation in Janus vdWs Nb3SeI7},
  author={Wang, Jiapeng and Yuan, Xiaojia and Fang, Yuqiang and Chen, Xinfeng and Zhong, Zhengbo and Lin, Shui and Qu, Jiafan and Fu, Jierui and Liu, Yue and Li, Zhipeng and others},
  journal={Small},
  volume={21},
  number={2},
  pages={2408628},
  year={2025},
  publisher={Wiley Online Library}
}

@article{Zhang2020Review,
   author = {Lei Zhang and Zhenjingfeng Yang and Tian Gong and Ruikun Pan and Huide Wang and Zhinan Guo and Han Zhang and Xiao Fu},
   doi = {10.1039/d0ta01999b},
   issn = {20507496},
   issue = {18},
   journal = {Journal of Materials Chemistry A},
   month = {5},
   pages = {8813-8830},
   publisher = {Royal Society of Chemistry},
   title = {Recent advances in emerging Janus two-dimensional materials: From fundamental physics to device applications},
   volume = {8},
   year = {2020}
}

@misc{CrysAlisPRO,
  title        = {{CrysAlis PRO}},
  year         = {2014},
  note         = {Agilent Technologies Ltd., Yarnton, Oxfordshire, England},
}
\bibliographystyle{rsc} 
\end{document}